\documentclass[conference]{IEEEtran}
\IEEEoverridecommandlockouts
\usepackage{amsmath,amssymb,amsfonts}
\usepackage{algorithmic}
\usepackage{cite}
\usepackage{graphicx}
\usepackage{listings}
\usepackage{textcomp}
\usepackage[color=green!40]{todonotes}
\usepackage{xcolor}

\def\BibTeX{{\rm B\kern-.05em{\sc i\kern-.025em b}\kern-.08em
    T\kern-.1667em\lower.7ex\hbox{E}\kern-.125emX}}

\begin{document}

\title{MTFS: Merkle-Tree-Based File System \\
}

\author{\IEEEauthorblockN{1\textsuperscript{st} Jia Kan}
  \IEEEauthorblockA{\textit{Department of Electrical and Electronic Engineering} \\
    \textit{Xi'an Jiaotong-Liverpool University}\\
    Suzhou, China \\
    Jia.Kan17@student.xjtlu.edu.cn} \and \IEEEauthorblockN{2\textsuperscript{nd}
    Kyeong Soo Kim}
  \IEEEauthorblockA{\textit{Department of Electrical and Electronic Engineering} \\
    \textit{Xi'an Jiaotong-Liverpool University}\\
    Suzhou, China \\
    Kyeongsoo.Kim@xjtlu.edu.cn}
}

\maketitle

\begin{abstract}
  The blockchain technology has been changing our daily lives since
  \textit{Bitcoin}---i.e., the first decentralized cryptocurrency---was invented
  and released as open-source software by an unidentified person or a group
  called Satoshi Nakamoto in 2009. Of many applications which can be implemented
  based on the blockchain, storage is an important one, a notable example of
  which is the InterPlanetary File System (IPFS). IPFS is a distributed web
  based on a peer-to-peer hypermedia protocol to make the web faster, safer, and
  more open and focuses on public accessible files. To provide a solution for
  private file storage in the blockchain way, in this paper we propose a
  Merkle-tree-based File System (MTFS). In MTFS, the blockchain is more than a
  trust machine; it is an abstract of a cluster system. Distributed random nodes
  form a tree network cluster without a central controller to provide a secure
  private storage service and faster message propagation. Advance proxy
  re-encryption algorithm is applied to guarantee secure file exchanges under
  permission. Merkle tree will make sure that the files are distributed among
  the service nodes in a balanced way. The proposed MTFS can be used not only
  for personal file storage and exchange but also for industry requiring mutual
  trust in file uploading and downloading in making contracts like insurances.
\end{abstract}


\section{Introduction}
The early Internet was created and used by academia mainly for research purpose:
The email was used to exchange ideas, and the file transfer protocol (FTP) was
used to exchange data like software packages and experimental results. Note that
centralized file servers then couldn't meet user demands all the time due to
their limitations in network bandwidth, input/output (I/O) speed, and/or storage
capacity. There are two major use cases for FTP: Public file distribution and
private access. In the following years, hypertext transfer protocol (HTTP)
replaced FTP for public file distribution as the world wide web (WWW) and
browser technology became widely popular. The concept of mirror was also used to
better serve users with higher speed from mirror servers geographically nearer
to users than the original server. Later the peer-to-peer (P2P) technology
brought another revolution, where a client downloading contents also serves
other clients with downloaded contents, named a peer.

On the other hand, the development of private data storage has been taking a
different path: FTP/secure FTP (SFTP) can be used for private data exchanges,
but, because they are not supported by Windows File Explorer, the common
internet file system (CIFS), a dialect of server message block (SMB) protocol,
is used in private file storage purpose more often, especially in enterprises
and commercial organizations. Due to the insufficient performance of CIFS/SMB,
more powerful and user-friendly commercial applications like Dropbox, Google
Drive, and Baidu Yun appeared, and similar services based on cloud also were
directly integrated into operating systems like Microsoft OneDrive and Apple
iCloud. BitTorrent Lab surprised us by introducing BitTorrent Sync (BTSync; now
Resilio Sync), which is a Dropbox-like application without the requirement of a
centralized server. Using existing P2P network, BTSync gets files synchronized
among personal devices.

Bitcoin\cite{ref_article1} based on blockchain and P2P technology. 
The blockchain technology brought us into a new era in private data storage as
well. InterPlanetary File System (IPFS) \cite{ref_article8} intends to build a
new distributed web.

Merkle-Tree-based File System (MTFS), which we propose in this paper, is to
provide a solution for private file storage based on P2P network and the
blockchain technology.  It uses the asymmetric cryptography including Proxy
Re-Encryption (PRE)\cite{ref_article10} technology to provide secure and
reliable private storage under permission. Random nodes form a cluster without a
central controller to provide a private storage service.

\section{Use Case}
The use of cloud services is common nowadays in enterprises, schools,
governments and even households for private files, and the proposed MTFS could
provide an alternative solution for private file storage/exchange based on the
blockchain technology.
We illustrate the use case of the blockchain-based private file storage/exchange
in the insurance industry as an example: People purchase insurance to cover
accidental loss in life. An insurance company asks a person to provide certain
information to insure him/her through risk evaluation. However, because those
submitted files are only kept on the insurance company's private server, it is
not possible for the insured to verify the submission of those files later
unless the insured kept the original reception as evidence. Blockchain-based
private file storage can provide mutual trust for both insurer and insured. When
the insured requests for insurance services, all supported materials are
provided in a restricted network space, which are immutable and reliable.

\section{Design}
Unlike public files which are usually shared among users with common interests,
like movies, albums, open courses and software packages, private files are
unique to each user and include confidential information (e.g., personal
documents, photos and videos). Hence security is critical to private file
storage in addition to speed and stability.

\subsection{Architecture}
In P2P, a peer is defined as both uploader and downloader, which combines the
role of a client and a server. A major concern for a P2P system is its
relatively lower performance; lots of peers are behind firewalls and often
connected with slow home connections like ADSL having much smaller uploading
bandwidth than downloading one.

In MTFS' design, a node is consisted of a batch of servers with professional
connection sitting in a data center. The host is assigned with public IP
address(es) and accessible open port(s) and the power supplied in 7x24. To build
a modern and efficient service, we must base it on fundamental infrastructure.


\subsection{Cryptography}
IPFS\cite{ref_article8} comes without built-in cryptography. To put a file in a
public place with the content encrypted safely, asymmetric cryptography
algorithm needs to be applied to make sure that only the private key owner can
decrypt the file content. In the application level, OpenPGP or GPG can be used
to encrypt and decrypt the file content.



Because nowadays mobile users take a large proportion, the limited bandwidth of
mobile connections and the limited power and storage space of mobile devices
should be considered in design. Consider the scenario of sending an encrypted
file to multiple users as an example. The following actions should be taken in
this case:
\begin{enumerate}
\item Collect each user's public key.
\item Encrypt the file with each user's public key.
\item Upload encrypted files to a server.
\end{enumerate}

Note that the step 2 costs too much computing power and local storage and that
the step 3 costs too much bandwidth. To address these issues, PRE\cite{ref_article11} 
is used to re-encrypt the decoding capsule without
modifying existing cipher texts.

\subsection{Broadcast network}
P2P network is a mandatory component of the blockchain. Among the three types of
P2P communication models (i.e., pair-wise, group-wise, and broadcast),
broadcasting is the most common requirement as every transaction or new block
discovery requires to be announced in the whole network as efficiently as
possible.

The broadcast network can be implemented with the gossip
protocol\cite{ref_article4} or DHT/KAD network\cite{ref_article3}. The gossip protocol is extremely reliable, but it
cannot ensure that the message can reach every corner of the network within a
fixed time. The propagation convergence could take lots of time.

We propose a tree network for broadcasting as show in Figure~1. In the tree
topology, a node joins the tree as a leaf one by one, and a protocol is set to
ensure the tree is constructed as balance as possible. In a tree network, the
longest distance between any two nodes will be less than 2 times of the tree
height. That means even in a binary tree topology, the message can be passed to
the farthest node with less than 2 times the tree height hops. It might be the
most efficient way for message broadcasting in logic.

\begin{figure}[htbp]
  \includegraphics[width=\linewidth]{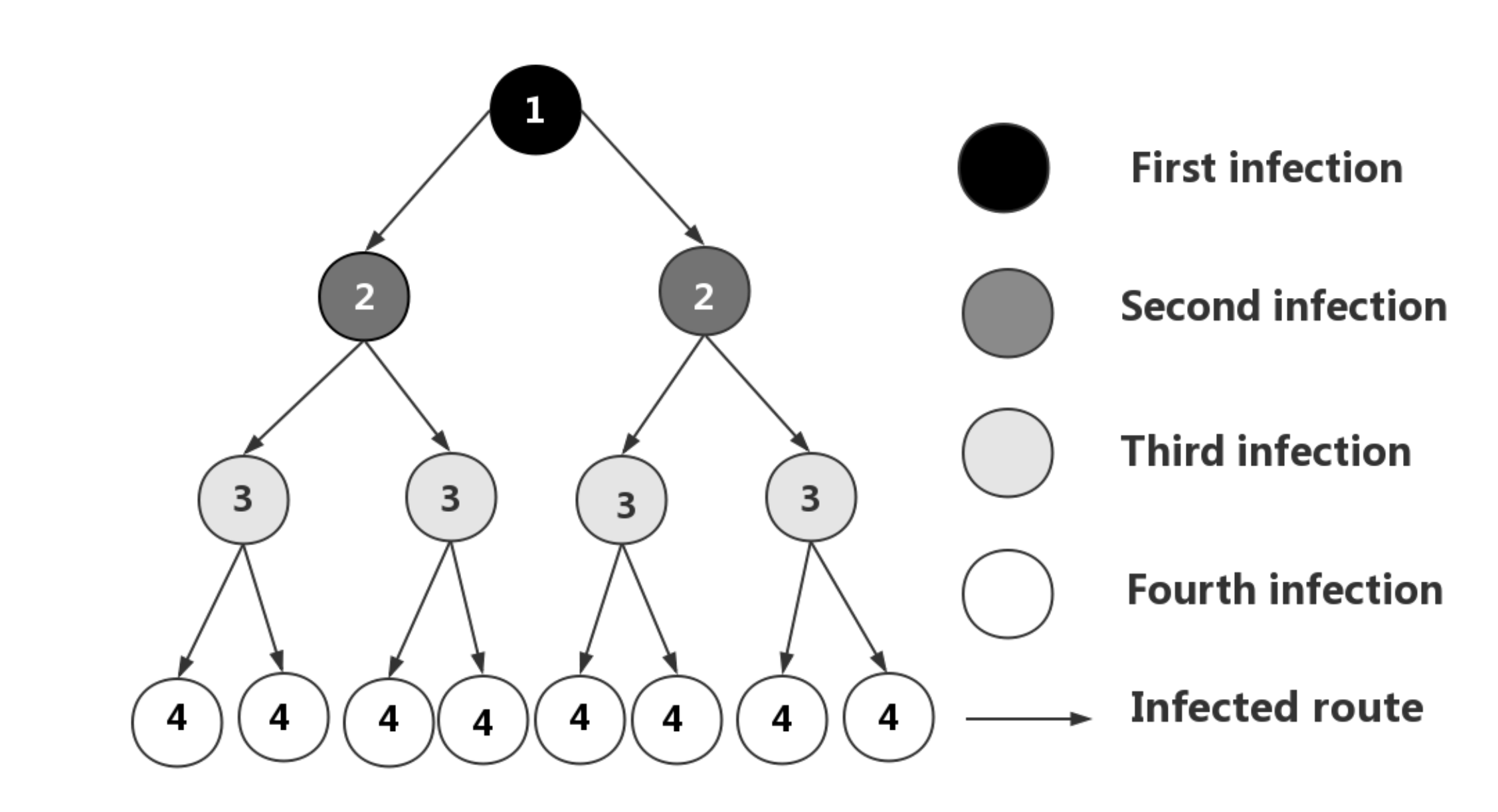}
  \caption{Tree-based broadcast network.}
\end{figure}

In the tree-based broadcast network, message could be initialized from any tree
branch node. A binary tree node will pass the message to its parent and two
children from the original node. A neighborhood receiving the message will
forward it to other neighborhoods except the one from which it received the
message.

Note that the tree network has also a disadvantage compared to the gossip
protocol: A single node failure can block the message propagation, because there
is only one way for a message to traverse. In the following, we discuss two
options to address this issue.

\subsubsection{Tree-based network with redundant nodes}
\label{sec:tree-based-network}
This option brings redundancy to a node by extending it to a cluster of multiple
nodes. For the experiment purpose, we consider a cluster of up to 3
nodes. Inside the cluster, the nodes interlink with one another.

\subsubsection{Tree-based network with redundant connections}
In this option, each node interlinks with not only the nearest neighbor but
also neighbors within several hops. This makes the tree network redundant when a
direct connection to the nearest neighbor fails; a message still can be passed
to the nearby nodes.

\subsection{Blockchain}
The last generation of P2P application like BitTorrent lacks the incentive
mechanism. Users can join the file downloading any time. In MTFS, it's required
for the contributor (i.e., miner) stay online as long as possible.

The blockchain technology will take a very important part in MTFS design. When a
user requests for certain file storage, a contract would be signed between the
user and the system. A subscription relationship will be set up, unless the user
decides to cancel or the node is about to quit.

\section{Implementation}
In the design of MTFS, Merkle tree\cite{ref_article2} is used to split encrypted
content to small pieces of objects (less or equal to 1 MB) and verify
the completeness. With all the existing design including\cite{ref_article11},
GraphChain\cite{ref_article6}, Tree network\cite{ref_article12}, we are
able to see the big picture of a private storage system now.

User will be able to start using MTFS by generating a public/private key
pair. To store a file on the network, the content needs to be encrypted with the
user's public key first. After the cipher text generated, a storage contract
between the user and a resource contributor will be signed, and the contract is
recorded in blockchain. MTFS will host the encrypted content for the user.

\subsection{Tree network construction}
The tree network is the backbone of MTFS system. It interlinks the distributed
system's resources, plays the role of communication infrastructure and
implements self governance. Each node in the network should be a server with
public IP address and open accessed port.

The first node in a tree network becomes root node, whose group identifier (ID)
is an empty string.
Any node in the tree can have up to two children nodes. After the establishment
of a connection, the parent node assigns the child node a group ID. Note that,
during the establishment of a connection with the parent node, the child node
itself can be a parent node for other nodes.

\subsubsection{Open branches}

The tree network is used for message broadcasting; the information of open
branches (Figure~2) is spreading and synchronized among nodes by message
broadcasting. Each node keeps a copy of information listing available branches
within the whole tree network.

Two message protocols are defined for the open branches information management:
AVAILABLE\_BRANCHES and DISCARDED\_BRANCHES. The message AVAILABLE\_BRANCHES is
used to announce new open connection points to the network that new nodes can
join the network. Another message DISCARDED\_BRANCHES indicates that current
branch already accepts a child node connection, so the node's information will
be removed from the list of globally available open branches.

When a child node managed to connect to a parent node, the parent node sends a
DISCARDED\_BRANCHES message to network due to its two available branches
taken. Meanwhile, the connecting child node will send a message
AVAILABLE\_BRANCHES after its parent node assigns it a group ID, telling the
network another two branches are open.


\subsubsection{Group Naming}
A node connects to a parent node; there might be two branches available for
attaching. A binary tree has a left and a right branch. When a branch is
attached, the parent will response a GROUP\_ID message to confirm the official
group ID of the newly-joined node. This action is to prevent a node to connect
to a tree branch which is already taken by other node. In such a case, the
parent will force to disconnect the duplicate connection.


In a binary tree, a group ID is represented in a binary format with a varied
length. The group ID of a root node is an empty string, while its left and right
child node take 0 and 1 respectively, as shown in Figure~1. Any hash hex string,
i.e., message digest like \textit{sha1} or \textit{md5}, can be easily converted
into a binary representation. As the tree network grows, the hash string can be
mapped to the outer nodes by the same prefix in a distributed manner.


\subsection{Discover of nearby nodes}
As the tree network grows, there are two things for each node to memorize: First
is the available branches, and second is all the nodes' host information. The
second requirement will be extremely expensive when the network size is large.
Instead of storing all the nodes' host information, it's cheaper to remember the
nearby neighbors' IP and port in memory. We can set a distance $k$, ask node to
remember the nodes within $k$ steps of hops. $k$ must be equal to or larger than
2; otherwise it's meaningless as any node already knows the direct neighbors'
information such as its parent and children.



\subsubsection{Recursive node discovery}
If the encrypted file size is more than 1 MB, the file is divided into
smaller objects by Merkle tree. Otherwise the file doesn't need to be divided,
and it can be directly uploaded to a node. By comparing the object hash with the
existing group IDs, user can recursively find the outermost node in the tree.


\subsection{Group Path}
\begin{figure}[htbp]
  \begin{center}
    \includegraphics[width=0.9\linewidth]{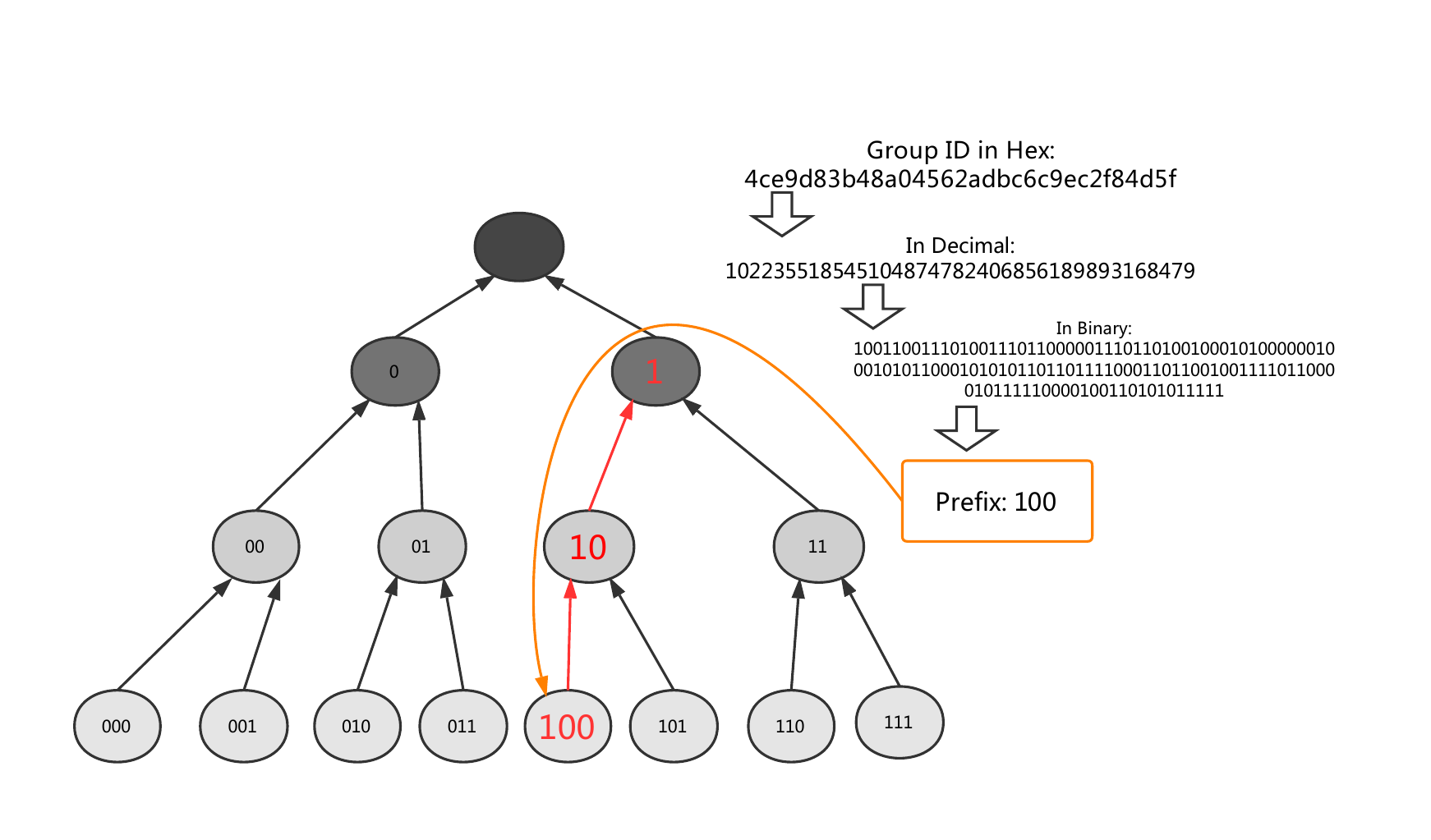}
    \caption{Tree-based network Naming and Group Path}
  \end{center}
\end{figure}

For a given hex hash string of node ID, it can be represented in a binary
format. Comparing it with the existing network, we can find nodes whose group
IDs are the prefix of the given hex hash string. Those nodes are defined as
Group Path as shown in Figure~2.

\subsection{Local encryption}
PRE is an asymmetric encryption algorithm. When user uploads a file to the
tree, it is required for the user to encrypt the files with the public
key. Besides the cipher text, another file named \textit{capsule} is
generated. The encrypted cipher text can be decrypted with user's private key
and the capsule.

Plain text is encrypted with PRE, resulting in a cipher text file and a capsule
file. The cipher text file size is usually as big as the plain text, while the
capsule file size is small.

\subsection{File and folder object}
The objects are generated from the encrypted content using Merkle tree. If the
encrypted content is less than 1 MB, only one object is created. In
this case, root hash equals to the encrypted content hash. Otherwise, a file
object named with root hash suffix "\_mt" is generated which contains Merkle
tree structure of file objects. Besides, there is capsule file named with root
hash suffix "\_capsule".

We can decode the file content with user's private key and the capsule, but the
file name is lost. The file name and size is stored in the folder object.

\subsubsection{Convention object size}
The convention object size of MTFS is 1 MB (i.e., 1024*1024 bytes). It
is the base size unit in MTFS. A small document file usually requires only one
unit of the object. For large-size photos or videos, however, more objects are
needed; in such a case, the allocation can be done in parallel for speedup.






\subsection{User and nodes}

\subsubsection{File uploading}
User's root folder object is created when the user uploads the first file or
creates a sub folder under the root folder.

The user is tended to choose the the edge node to store new objects; we can
imagine the network like a city, where its central area is always busy with
traffic jam, while its country side far from the central area has less traffic
and enough land.

The user's files, once encrypted and split into objects, are uploaded to the
tree by comparing group ID and object hash and finally stored into MTFS. Then,
the user's root folder information is uploaded to MTFS as an object. The folder
object is updated following the same process as the file object.

As the folder and file objects are saved into the tree, a contract is set up
between the user and the nodes. The root folder object's hash is committed into
blockchain as a transaction. Besides the root hash, the transaction also
includes the group ID information.

\subsubsection{File retrieving}
Inversely, when user browses file in MTFS, the first step is to look for the
root folder information on the nodes. The user's root folder object and location
information (i.e., group ID of storage) is committed as the transaction of
blockchain. With object ID and group ID, it's easy to retrieve the content of
the root folder object. If the folder content is larger than 1 MB, the
user will follow the procedure to gather all the folder objects and decrypt
them.



\subsubsection{Folder object schema}
User checks the group path on the tree, where there should be a node hosting
the user's information (assume the network keeps 3 replications).

A folder object is different from a file object. A folder object contains the
encrypted folder data. Data are represented in JSON format. Entities in folder
data link to other file object and folder object. File object size indicates the
file size, and folder object size indicates the total file size under the
folder. This information is useful to calculate how much storage consumed by the
user.

If the file size is larger than 1 MB, use the
Merkle tree root hash; otherwise use the hash of the file content.



\subsection{File exchange}
File exchange is similar to sending an email to another user. A sender should
know the address of a receiver, i.e., the public key. The file sending action is
a blockchain transaction operation; the transaction shows who and which file on
MTFS will be added into the receiver's folder.

MTFS is a secure system with permission. In the operation of file sending, no one except the
owner has the permission to modify the folder information. However, user may interact with MTFS from a mobile device. It would be too
expensive to do operations like downloading, decrypting, modifying, encrypting,
and uploading in this case. Note that, if the user wants to modify the content,
this series of operations has to be done. What if the user does not modify
the content, was it possible to reduce the operation steps? PRE
is introduced for this case.

With PRE, the file owner can encrypt
the file content into a cipher text and a capsule. The cipher text is as long as
the content, but the capsule is usually very short. The capsule can be
re-encrypted with sender's private key and receiver's public key.
Once the receiver accepts the re-encrypt capsule granted by sender, the
receiver can easily decrypt the cipher text with the new capsule and his or her
own private key.


\subsection{Replication and verification}
After the file is uploaded to the system, the file must replicate itself for
several copies on the group path. Because the objects on the node is protected
with authentication, the other nodes on the group path do not have the
permission to pull the objects. So, it is required that the node pushes objects
to the replicated nodes. The verification of the storage needs to be performed periodically, as the
object content may be damaged, lost, or even the node may try to cheat by
claiming that they have the objects but actually not.

\section{Conclusions}
In this paper, we have proposed MTFS and discussed its many aspects from the
technology requirements to the design and implementation with major focus on
cryptography, blockchain, network and storage schema.

Based on a novel design we building a secure
private file storage system using existing technologies: A tree network is used as a backbone network for
storage with high performance broadcasting, where only servers with public IP
addresses are allowed to join as nodes to avoid the issues related with firewall
and gateway NAT in achieving higher performance. Also PRE encryption is
introduced to re-encrypt the decoded capsule without modifying existing cipher
texts. Base on these component technologies, the schema of storage is designed,
and MTFS can provide a high-performance solution for a secure private file
storage service based on the blockchain technology.





\vspace{12pt}


\begin{thebibliography}{00}

\bibitem{ref_article1} Nakamoto, S. (2008). Bitcoin: A peer-to-peer electronic
  cash system.

\bibitem{ref_article2} Merkle, R. C. (1987, August). A digital signature based
  on a conventional encryption function. In Conference on the theory and
  application of cryptographic techniques (pp. 369-378). Springer, Berlin,
  Heidelberg.

\bibitem{ref_article3} Maymounkov, P., \& Mazieres, D. (2002, March). Kademlia:
  A peer-to-peer information system based on the xor metric. In International
  Workshop on Peer-to-Peer Systems (pp. 53-65). Springer, Berlin, Heidelberg.

\bibitem{ref_article4} Demers, A., Greene, D., Hauser, C., Irish, W., Larson,
  J., Shenker, S., ... \& Terry, D. (1987, December). Epidemic algorithms for
  replicated database maintenance. In Proceedings of the sixth annual ACM
  Symposium on Principles of distributed computing (pp. 1-12). ACM.


\bibitem{ref_article6} Kan, J., Chen, S., \& Huang, X. (2018). \textit{Improve
    Blockchain Performance using Graph Data Structure and Parallel Mining.}
  arXiv preprint arXiv:1808.10810.


\bibitem{ref_article7} Bi, W., Yang, H., \& Zheng, M. (2018). An Accelerated
  Method for Message Propagation in Blockchain Networks. arXiv preprint
  arXiv:1809.00455.

\bibitem{ref_article8} Benet, J. (2014). IPFS-content addressed, versioned, P2P
  file system. arXiv preprint arXiv:1407.3561.


\bibitem{ref_article10} Wikipedia, (2017). Proxy re-encryption - Wikipedia, the
  free encyclopedia.  Available at:
  https://en.wikipedia.org/wiki/Proxy\_re-encryption

\bibitem{ref_article11} David, N. (2018). Umbral: a threshold proxy
  re-encryption scheme. Available at:
  https://github.com/nucypher/umbral-doc/blob/master/umbral-doc.pdf

\bibitem{ref_article12} Kan, J., Zou, L., Bella, L., \& Huang,
  X. (2018). \textit{Boost Blockchain Broadcast Propagation with Tree Routing.}
  arXiv preprint arXiv:1810.12795.


\end{thebibliography}
\end{document}